%% file: main_revised.tex
\documentclass[10pt]{article}
\input{hicss-packages}
\usepackage{amsmath,amsfonts,amssymb,amsthm}
\usepackage{algorithmic}
\usepackage{algorithm}
\usepackage{array}
\usepackage{textcomp}
\usepackage{stfloats}
\usepackage{verbatim}
\usepackage{tabularx}
\usepackage{multirow}
\usepackage{adjustbox}
\usepackage{upgreek}      
\usepackage{diagbox}      
\usepackage{xcolor}
\usepackage{pifont}       
\usepackage{stmaryrd}     
\usepackage{booktabs}     
\usepackage{makecell}
\usepackage[utf8]{inputenc}
\usepackage{hyperref}
\hypersetup{
  colorlinks=false,
  hidelinks
}

\renewcommand{\arraystretch}{1}

\def\BibTeX{{\rm B\kern-.05em{\sc i\kern-.025em b}\kern-.08em
    T\kern-.1667em\lower.7ex\hbox{E}\kern-.125emX}}

\title{Machine Learning-Based State Estimation for an Actual Transmission System Using Field PMU Data}

\author{Shiva Moshtagh \\
 School of Electrical, Computer,\\ and Energy Engineering,\\
 Arizona State University \\
 {\underline{\href{mailto:smoshta1@asu.edu}{smoshta1@asu.edu}}}
 \And 
 Nihar Thakkar\thanks{\quad N. Thakkar is now with Google.} \\
 School of Computing\\ and Augmented Intelligence,\\
 Arizona State University \\
 {\underline{\href{mailto:njthakka@asu.edu}{njthakka@asu.edu}}}
 \And 
 Anamitra Pal \\
 School of Electrical, Computer,\\ and Energy Engineering,\\
 Arizona State University \\
 {\underline{\href{mailto:Anamitra.Pal@asu.edu}{Anamitra.Pal@asu.edu}}}
 \And 
 Evangelos Farantatos \\
 Electric Power Research\\ Institute (EPRI) \\
 {\underline{\href{mailto:efarantatos@epri.com}{efarantatos@epri.com}}}}

\date{}

\begin{document}
\begingroup
\allowdisplaybreaks

\maketitle

\begin{abstract}
Time-synchronized state estimation (SE) plays a critical role in ensuring real-time situational awareness in modern power systems. However, achieving full system observability using phasor measurement units (PMUs) is often impractical due to cost and deployment constraints. Moreover, SE operation at PMU timescales imposes stringent requirements on latency, robustness, and reliability that are difficult to satisfy using conventional iterative hybrid SE techniques under incomplete observability by PMUs. This paper evaluates the feasibility of deploying deep neural networks for 
PMU-timescale, time-synchronized SE in real-world PMU-unobservable transmission systems using actual data from a US power utility. Key contributions include a systematic assessment of estimation accuracy, scalability, and computational performance under realistic operating conditions.
\end{abstract}

\subsubsection*{Keywords:}

Machine Learning, Phasor Measurement Units (PMUs), State Estimation (SE), Unobservability.


\section{Introduction}
Modern power systems are becoming increasingly complex due to the integration of renewable energy sources, dynamic loads, and advanced control technologies \parencite{yusuf2025review}.
These factors introduce new operational challenges that demand reliable and real-time visibility of system states to ensure secure and efficient operation \parencite{ali2025power}.
State estimation (SE) serves as a critical function within the energy management system (EMS), providing operators with the most accurate snapshot of network conditions for decision-making and control \parencite{cheng2024survey}.

Conventional weighted least squares (WLS)-based SE methods have been used in transmission system EMS for decades.
They rely primarily on supervisory control and data acquisition (SCADA) measurements, with phasor measurement unit (PMU) data playing a smaller role due to \textit{deployment}, \textit{communication}, and \textit{cost} constraints \parencite{phadke2017synchronized}.
These constraints have resulted in most transmission systems lacking complete observability by PMUs, particularly at lower voltage levels \parencite{varghese2024deep}.
As such, despite PMUs providing time-synchronized measurements at high speeds, there has been limited adoption of PMUs as the primary data source for SE \parencite{zhang2025graph}.
Moreover, conventional hybrid SE techniques are not well-suited for execution at PMU reporting rates due to iterative optimization-based estimation \parencite{cheng2024survey}.
From an operational standpoint, achieving accurate and computationally efficient SE under partial observability remains one of the key unresolved challenges for transmission system operators, particularly in networks with sporadic PMU placement and fixed infrastructure constraints.

PMU deployment is primarily driven by protection functions, stability monitoring needs, and communication infrastructure availability, rather than by SE observability requirements. As a result, many transmission networks exhibit persistent, topology-dependent observability gaps that remain unchanged over long operational periods.
To address these gaps, there has been
growing industrial and academic interest in developing machine learning (ML)-based SE frameworks for unobservable power systems.
Early shallow learning approaches \parencite{kumari2017data,tian2021neural} demonstrated feasibility but suffered from limited scalability and poor generalization to unseen operating conditions (OCs).
Subsequent deep learning-based estimators evaluated on benchmark IEEE test systems such as the 14-bus and 118-bus networks \parencite{wang2020physics,carmichael2022application,mestav2019learning} established the theoretical advantages of deep neural network (DNNs)-based-SE (DNN-SE) under partial observability, leveraging the ability of DNNs to directly map available measurements to system states for fast, non-iterative inference.
However, these works relied exclusively on synthetic power flow (PF)-derived measurements, omitting practical issues such as bad data found in the field, sensor-level anomalies, and channel mapping inconsistencies.
More recent works targeting practical implementation challenges \parencite{azimian2022state,varghese2024deep,moshtagh2023time,moshtagh2025topology} have addressed topology changes, PMU unobservability, and bad data handling, yet validation in these works still relies on simulated PMU measurements derived from PF solutions rather than measurements streamed from field-installed PMUs.


Unlike the majority of existing simulation-oriented studies, this paper focuses on the application and validation of ML-based SE using real-world PMU and historical SCADA data from an actual transmission system.
To the best of our knowledge, \textit{this is the first study to systematically evaluate a DNN-based time-synchronized SE framework using field PMU data}.
Rather than introducing a new SE algorithm, this work serves as a validation study, focusing on the practical applicability, scalability, accuracy, and computational efficiency of DNN-based SE 
under realistic OCs.

\section{ML-Based SE Framework}
\label{section:DNN-SE}
The objective of ML-based SE is to enable fast and reliable estimation of system states by learning a direct mapping between available measurements and network states.
In contrast to conventional model-based estimators that rely on explicit system equations and iterative optimization, data-driven approaches leverage historical operational data to infer this mapping directly \parencite{venzke2020learning}.
In this work, DNNs are employed due to their strong approximation capability and suitability for high-dimensional nonlinear regression problems \parencite{sonoda2017neural}. Once trained, the resulting estimator enables non-iterative, low-latency SE using time-synchronized PMU data,
making it well suited for real-time operational environments.

\subsection{Basics of DNN-based SE (DNN-SE)}
From a probabilistic perspective, DNN-SE can be interpreted within a \textit{Bayesian} estimation framework.
Unlike conventional WLS methods that minimize the \textit{modeling error},
the Bayesian approach seeks to minimize the \textit{expected estimation error} by computing the minimum mean-squared error (MMSE) estimate of the system state.
In this framework, both the state variables \(x\) and the measurements \(z\) are treated as random variables, and the goal is to estimate the state by minimizing the MMSE.
Under this formulation, the optimal state estimate \( \hat{x}^*(z) \) that minimizes the mean-squared estimation error is given by the conditional expectation $\mathbb{E}[x | z]$ as follows:
\begin{equation}
    \min_{\hat{x}(\cdot)} \mathbb{E} \left( \| x - \hat{x}(z) \|^2 \right) \Rightarrow \hat{x}^*(z) = \mathbb{E}[x | z]
\end{equation}

The MMSE estimator directly minimizes the expected \textit{estimation error} (unlike traditional estimators that minimize the \textit{modeling error}), which circumvents the need for complete observability. The minimization is achieved by leveraging prior knowledge about the system along with the observed data to compute a posterior distribution $p(x | z)$ for the state variables \parencite{mestav2019bayesian}.
The MMSE estimate can then be obtained from the posterior distribution as:
\begin{equation}
    \hat{x}^*(z) = \mathbb{E}[x | z] = \int x p(x | z) dx
\end{equation}

However, it is difficult to analytically calculate the above-mentioned posterior distribution and/or perform the resulting integration when the system has limited observability.
In such a scenario, deep learning models such as DNNs come to the rescue as they have excellent approximation capabilities \parencite{sonoda2017neural}.
That is, the deep learning models are employed to approximate the MMSE estimator.

The MMSE estimator possesses two foundational properties relevant to operational SE \parencite{mendel1995lessons}. First, it is \textit{unbiased}, i.e., $\mathbb{E}[\hat{x}^*(z)] = \mathbb{E}[x]$, ensuring
consistency with the underlying state distribution. Second, among all estimators (linear or nonlinear), it achieves the \textit{minimum error variance}, given by:
\begin{equation}
J^*(\tilde{x}) = \mathbb{E}[x^T x \mid z] - \mathbb{E}[x^T \mid z]\,\mathbb{E}[x \mid z]
\end{equation}
which corresponds to the conditional variance of $x$ given $z$. Together, these properties establish $\mathbb{E}[x|z]$ as the theoretically optimal estimator target, motivating its use as the learning objective in our DNN-based framework. While analytical MMSE approaches such as a Gaussian mixture model-based estimator \parencite{bilil2018mmse} are effective for small systems, they become intractable at the scale of real transmission networks, motivating the DNN-based approximation employed here.

The proposed DNN-SE formulation does not require the system to be fully observable in the classical sense, as the estimator leverages statistical dependencies learned from historical SCADA data to infer unobserved state variables at PMU timescales.
Moreover, by relying exclusively on time-synchronized PMU measurements during online inference, the approach avoids the synchronization challenges associated with hybrid SCADA/PMU SE \parencite{zhao2018robust,darmis2023survey}.
The framework also exhibits robustness to moderate errors and noise present in the historical SCADA data used during the training phase.
More details about the performance of ML-based SE, including analytical performance guarantees can be found in \parencite{azimian2025application,azimian2023analytical}.

\subsection{Offline Training and Online Inference Workflow}
\label{Sec2.2}
The proposed DNN-SE framework follows a two-stage workflow consisting of an offline training and an online inference phase, as illustrated in Fig.~\ref{fig:diagram}. The key components of each phase are described below.

\begin{figure*}
    \centering
    \includegraphics[width=1\linewidth]{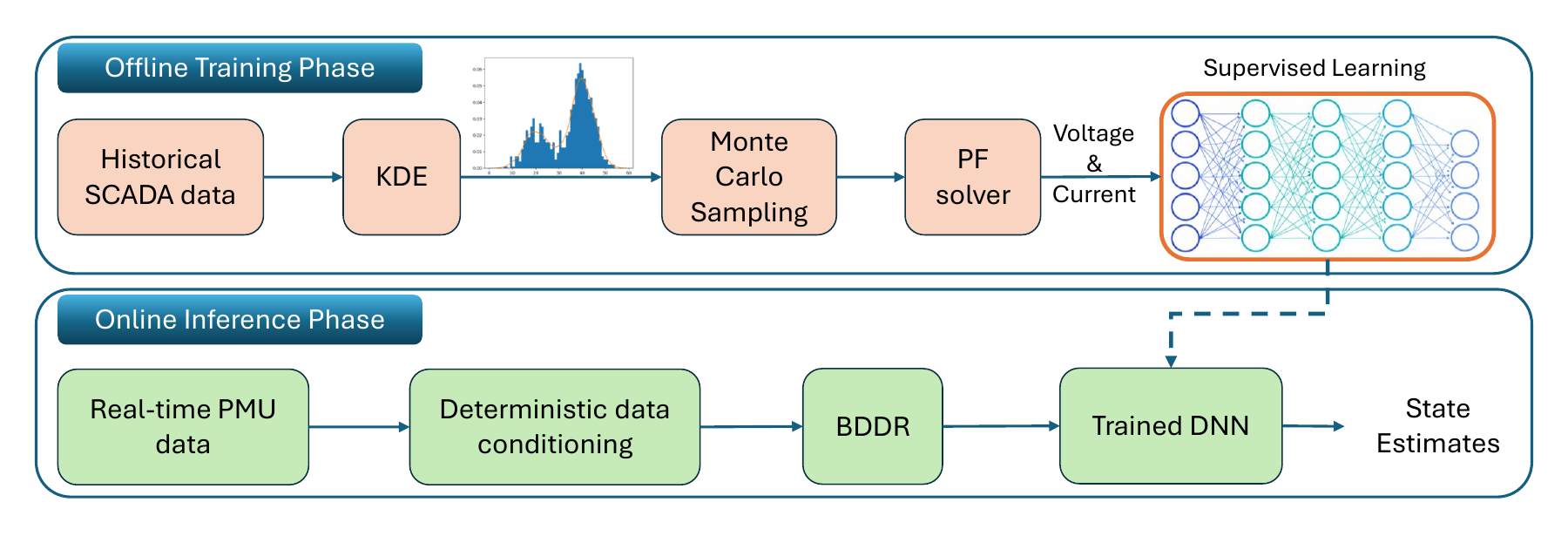}
    \caption{Offline training and online inference workflow of the DNN-SE framework. During the offline phase, historical SCADA snapshots are used to statistically characterize load behavior and generate additional OCs through Monte Carlo sampling. These sampled load profiles are processed using a PF solver to obtain reference system states for supervised learning. During online operation, the trained DNN receives time-synchronized PMU measurements that have undergone preprocessing and produces state estimates at PMU reporting rates without requiring SCADA data or iterative computations.}
    \label{fig:diagram}
\end{figure*}

\textit{Offline training phase:}
Historical SCADA data collected from the transmission system are used exclusively during the offline phase to generate training data for the DNN-SE.
Due to limited number of historical SCADA snapshots, a data augmentation strategy is employed to generate a sufficiently rich set of OCs.
Specifically, historical SCADA load measurements are statistically characterized using Kernel density estimation (KDE) to model the empirical behavior of individual loads.
These distributions are then sampled to generate additional load realizations, which serve as inputs to a PF solver based on the available network model. The resulting voltage magnitude and angle profiles constitute reference system states used for DNN training.
For each OC, the corresponding PMU measurements (voltage and current phasors) at installed PMU locations are extracted and used as inputs to the DNN, while the PF-derived system states are used as target outputs. In this manner, the DNN is trained to \textit{learn a mapping from limited PMU measurements to the full system state}.
It is emphasized that the solver-generated states are treated as best-available reference labels rather than exact ground truth, and that SCADA measurements are not required during online operation.

The reliance on offline training is operationally practical: training is performed once and fully amortized at deployment, while inference requires only a single forward pass through the DNN. The use of historical SCADA snapshots is feasible because transmission utilities maintain extensive multi-year SCADA archives, and the KDE-based characterization captures empirical load distributions, which mitigates sensitivity to individual SCADA measurement errors.

\textit{Online inference phase:}
During online operation, the trained DNN receives PMU data as inputs and quickly produces estimates of the full system state through a single forward pass of the network.
No SCADA measurements, PF calculations, or iterative optimization procedures are required during this phase.
Because all computationally intensive processing is confined to the offline training stage, the online execution exhibits deterministic latency and is suitable for PMU-timescale SE in operational EMS environments.

Prior to DNN inference, incoming PMU measurements undergo 
data conditioning steps, including data cleaning, reference frame alignment, and phase angle unwrapping, to ensure structural consistency with the training data and numerical stability. These operations are applied uniformly to all PMU inputs. 
To mitigate the adverse impact of bad data commonly observed in field PMU measurements, a bad data detection and replacement (BDDR) scheme is subsequently employed as a preprocessing step during online operation.
The adopted BDDR approach is based on a statistical hypothesis testing framework that employs 
the Wald Test, and is designed to identify and correct outlier measurements prior to DNN inference \parencite{mestav2019bayesian,liu2018multichannel}.
The detection scheme operates at PMU timescales and is well-suited for integration with ML-based SE frameworks.
The identified bad data points are replaced using appropriate data from the offline training phase, ensuring that the DNN receives inputs consistent with its training distribution (see \parencite{varghese2024deep} for more theoretical details about the replacement approach).

\section{Real-World Transmission System Setup}\label{section:TVA}
The case study in this work is based on an actual large-scale transmission system located in the US. Due to data confidentiality and proprietary constraints, the system is anonymized and is referred to as System $\mathrm{S}$ throughout the paper. The network model and associated operational data are described below in detail.

\subsection{Test System Description}
System $\mathrm{S}$ represents a multi-area transmission network comprising thousands of buses spanning voltage levels from $69~\text{kV}$ to $500~\text{kV}$.
The majority of the system infrastructure is concentrated at the $161~\text{kV}$ level, which constitutes the largest portion of the network in terms of both number of buses and number of branches. Higher voltage levels form the backbone of the system, while lower voltage levels provide regional interconnections.

SCADA snapshots of this system were provided over a six-month period.
Due to routine operational activities such as maintenance outages, switching actions, and seasonal operational adjustments, the system exhibited variations in its bus-branch configuration over time.
As a result, the available historical SCADA data corresponded to multiple network topologies rather than a single fixed configuration. To systematically characterize these variations, topology clustering\footnote{Based on branch-difference metrics and K-means clustering.} was performed by comparing bus and branch connectivity across the SCADA snapshots.
Although a variety of topologies were identified from the SCADA snapshots, the analysis done here
is based on a topology that was dominant between mid-July through mid-August.
This topology was selected because it aligns with the availability of actual PMU measurements and their corresponding metadata used for model testing.

\subsection{PMU Locations}
PMUs are installed across more than 100 substations in System $\mathrm{S}$ and are distributed across multiple transmission voltage levels, with the highest concentration being at the $161~\mathrm{kV}$ level.
Many substations host multiple PMUs monitoring different voltage levels simultaneously.
Of the $1{,}493$ transmission buses ($\geq$$69~\text{kV}$) considered in this study, approximately $133$ buses are directly monitored by PMUs, corresponding to a direct observability of only $\sim$$9\%$.
Due to this localized nature of PMU placement, the overall system remains only partially observable with none of the transmission voltage levels being fully observed by PMUs.
This insufficient PMU coverage limits the applicability of conventional PMU-based linear state estimators in operational settings \parencite{mishra2023algebraic}.

\subsection{Data Characteristics}

Three primary data sources
were utilized in this study: historical SCADA snapshots, actual PMU measurements, and supplementary metadata.
Historical SCADA snapshots were provided at half-hourly intervals in the form of PSS/E \texttt{.raw} files for
a six-month period from July 1 to December 31.
Two snapshots were recorded per day at 2:00 PM and 2:30 PM local time, resulting in approximately 366 raw files.

The field PMU data included positive-sequence measurements of voltage and current phasors and local frequency. PMU data used in this analysis was obtained at a resolution of $1$~frame per second over a one-hour interval (2:00~PM to 3:00~PM local time) for the 15th of August.
Supplementary metadata were used to map PMU channel identifiers to corresponding PSS/E bus and branch elements and to support topology clustering.

\section{Case Studies and Performance Evaluation}\label{section:results}

\subsection{Training Dataset Generation}
\label{Sec4.1}
The training dataset for the DNN-SE framework is generated using historical SCADA snapshots following the offline workflow described in Section~\ref{section:DNN-SE} and illustrated in Fig.~\ref{fig:diagram}.
As discussed in Section~\ref{section:TVA}, multiple network topologies were observed over the six-month study period.
Consistent with the topology selection process described earlier, training data generation efforts focused on the subset of snapshots corresponding to the period for which PMU data was available. This resulted in approximately 70 cases (two snapshots per day from July 12 to August 15), and the identified topology is henceforth referred to as topology $\mathrm{T}$.
Given this limited number of historical operating snapshots, a data augmentation strategy was employed to generate a sufficiently large and diverse training dataset based on KDE and random sampling as explained in Section \ref{section:DNN-SE}.
Fig.~\ref{fig:kde_loads} illustrates representative density estimates fitted to historical load data for ten  load buses.

\begin{figure}
    \centering
    \includegraphics[width=1\linewidth]{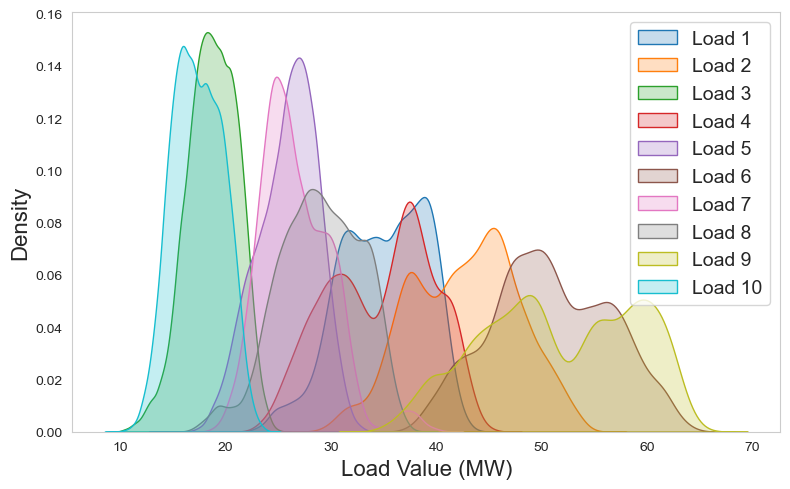}
    \caption{Distribution of 10 random loads of System $\mathrm{S}$ fitted by KDE on historical SCADA data. The fitted distributions are used to generate additional OCs through Monte Carlo sampling during the offline training phase.}
    \label{fig:kde_loads}
\end{figure}

For each sampled load realization, a PF analysis was performed in PSS/E using the network model corresponding to topology $\mathrm{T}$ to obtain the associated system state. This process was repeated for 
$25{,}000$ instances, resulting in a comprehensive set of 
OCs that capture realistic variability in system loading. The resulting PF solutions provide reference voltage magnitudes and angles for all buses, which serve as target outputs for supervised learning.
For each OC, the corresponding PMU measurements were constructed by extracting the subset of voltage and current phasors available at the installed PMU locations. These PMU measurements serve as inputs to the DNN during training, while the PF-derived system states serve as reference labels for the outputs. It is emphasized that actual PMU measurements are not used during the training phase and are reserved exclusively for online evaluation, as explained in the following sections.

\subsection{Experiment Setup and DNN-SE Architecture}
The system states targeted for estimation consist of the voltage magnitudes and 
angles at all buses operating at $69~\text{kV}$ and above.
For topology $\mathrm{T}$,
the total number of state variables is 
$2$,$986$, which corresponds
to voltage magnitudes and angles of $1$,$493$ buses.
Two DNN-SE models are implemented and evaluated.
The first model, 
\textit{DNN-SE$_{V}$}, utilizes only voltage phasor measurements collected at PMU locations.
The second model, 
\textit{DNN-SE$_{VI}$}, incorporates both voltage and current phasor measurements from PMUs.
The objective of this comparison is to assess the impact of incorporating additional measurement features on estimation performance.
Both DNN-SE models are trained using the dataset described in Section \ref{Sec4.1}.

The DNN architecture adopted in this study follows a sequential feed-forward structure with fully connected hidden layers and is implemented using Keras/TensorFlow.
The hyperparameters
for both DNN-SE models are summarized in Table~\ref{tab:hyperparameters}.
All simulations were conducted on a high-performance computer equipped with 256 GB of RAM, an Intel Xeon 6246R CPU operating at 3.40 GHz, and an Nvidia Quadro RTX 5000 with 16 GB of GPU memory.

\begin{table}[t]
\centering
\caption{Hyperparameters for DNN Training}
\label{tab:hyperparameters}
\renewcommand{\arraystretch}{1.15}
\setlength{\tabcolsep}{3pt}
\begin{tabularx}{\columnwidth}{lXX}
\toprule
\textbf{Hyperparameter} & \textbf{DNN-SE$_{V}$} & \textbf{DNN-SE$_{VI}$} \\
\midrule
Input Dimension & 266 & 746 \\
Output Dimension & 2986 & 2986 \\
Number of Hidden Layers & 4 & 4 \\
Neurons per Layer & 1500 & 2000 \\
Hidden Layer Activation & Rectified Linear Unit (ReLU) & Rectified Linear Unit (ReLU) \\
Output Layer Activation & Linear & Linear \\
Loss Function & Mean Squared Error (MSE) & Mean Squared Error (MSE) \\
Optimizer & Adam & Adam \\
Learning Rate & 0.001 & 0.001 \\
Dropout Rate & 0.3 & 0.3 \\
Batch Size & 300 & 300 \\
Number of Epochs & 300 & 300 \\
Regularization & Batch Nor- malization & Batch Nor- malization \\
\bottomrule
\end{tabularx}
\end{table}

\subsection{PMU Dataset and Online Evaluation Processing}\label{section:online}
\subsubsection{Deterministic PMU preprocessing.}\par
Once the DNN training is complete, model evaluation is performed using actual PMU measurements. Prior to being used as inputs to the DNN, the PMU data undergo a series of preprocessing steps designed to ensure consistency with the training data and to mitigate measurement artifacts commonly encountered in field deployments.
First, extensive data cleaning is performed to remove invalid entries, including missing values (NaNs) and stale/zero data, and to ensure consistent mapping between PMU channels and corresponding network elements. Duplicate PMU measurements associated with dual-use line relay configurations are then identified and removed to avoid redundant inputs.
After these steps, the final PMU input feature set is constructed, with dimensionality consistent with the input specifications reported in Table~\ref{tab:hyperparameters}.

Next, all voltage and current phase angle measurements are referenced to the system slack bus to maintain consistency with the reference frame used in the PF-based training data. This alignment is essential to ensure that the angular states estimated by the DNN are comparable across training and evaluation phases.
Finally, phase angle unwrapping is applied to eliminate artificial discontinuities caused by angle wrap-around effects. Without this correction, abrupt jumps may appear in the PMU data adversely impacting the accuracy of the DNN inference process.
Note that these preprocessing steps are 
distinct from the BDDR procedure 
described in Section \ref{Sec2.2}.

The PMU measurements are available at a resolution of $1$ frame per second over the interval from 2:00~PM to 3:00~PM local time on August~15. However, the historical SCADA snapshots used for training are available only at 2:00~PM and 2:30~PM. To ensure temporal consistency between training and evaluation conditions, the DNN models are evaluated using PMU data collected between 2:25~PM and 2:30~PM.
This five-minute evaluation window yields approximately 300 PMU samples and is sufficiently close to the 2:30~PM SCADA snapshot used during training data synthesis. The selected time window also enables the use of preceding PMU data for input normalization, as described next.

\subsubsection{Prior window normalization.}\par
Normalization of input features is essential for stable DNN training and inference. 
While the training inputs are usually normalized using standard scaling based on the statistics of the training dataset, directly applying the same normalization parameters to field PMU data may be inappropriate for our case as the PMU measurements do not necessarily lie within the range of the synthetic training data.
This happens because
the training inputs are synthetically generated from KDE-sampled SCADA snapshots processed through a PF solver, while the test inputs are streamed from field PMUs that carry sensor-level offsets, instrument transformer calibration factors \parencite{varghese2026system,varghese2026constrained}, and operational conditions not represented in the synthetic training distribution. Directly applying training-set statistics to field PMU inputs would therefore map them into a range the DNN never encountered during training, degrading estimation accuracy. Prior window normalization mitigates this by re-scaling incoming PMU inputs using their own recent statistics, effectively aligning the normalized input distribution seen at test time with the normalized distribution the DNN was trained on.
Note that this
prior window normalization strategy is only employed during online evaluation.

\subsubsection{Evaluation strategies.}\label{section:eval-strategies}\par
The evaluation conducted using field PMU data introduces a challenge not present in simulation-based SE studies: the absence of ground-truth state labels for the field data. To address this, two complementary strategies are adopted in this sub-section for the field data-based evaluation (\textit{common states} and \textit{SCADA snapshots}), while a third strategy based on a \textit{held-out synthetic test set} is used in the next sub-section to validate the DNN architecture itself in a setting where ground truth is exactly known.

\textit{Evaluation using common states:} A subset of system states corresponds to buses equipped with PMUs, for which voltage magnitude and angle measurements are directly available as DNN inputs.
These states, referred to as \textit{common states}, appear both in the input as well as in the output of the DNN. For these buses, the PMU measurements are treated as reference values, and the estimated states produced by the DNN are compared against the corresponding input measurements to assess estimation consistency and accuracy.
A representative example of this comparison for a PMU-equipped bus is
shown in Fig.~\ref{fig:PMU_buses} for voltage magnitudes and phase angles, respectively.

\begin{figure}[ht]
    \centering
    \includegraphics[width=1\linewidth]{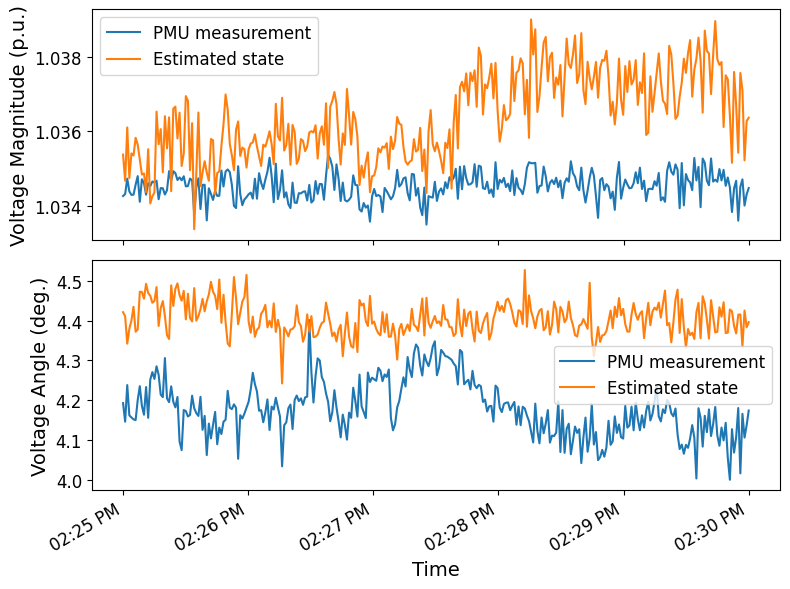}
    \caption{PMU measurement comparison with estimated state by DNN-SE$_{VI}$ for a bus equipped with PMU. The closeness of the plots are an indication of the high accuracy of this DNN-SE model.}
    \label{fig:PMU_buses}
\end{figure}

\textit{Evaluation using historical SCADA snapshots:}
In this strategy, the state estimates obtained over the evaluation window from the DNN are averaged over time. Then, these averaged estimates are compared with the system state values for every bus obtained from the 2:30~PM SCADA snapshot which is closest to the evaluation window (of 2:25~PM-2:30~PM).
Although SCADA measurements are not time-synchronized and do not represent exact ground truth, they are expected to provide reasonable approximations of the system state at the corresponding time.
A visual illustration of this comparison across all buses is presented in Fig.~\ref{fig:SE_per-bus} for both voltage magnitudes and phase angles.
It is clear from the figure that the estimates are very close together.

\begin{figure}[ht]
    \centering
    \includegraphics[width=1\linewidth]{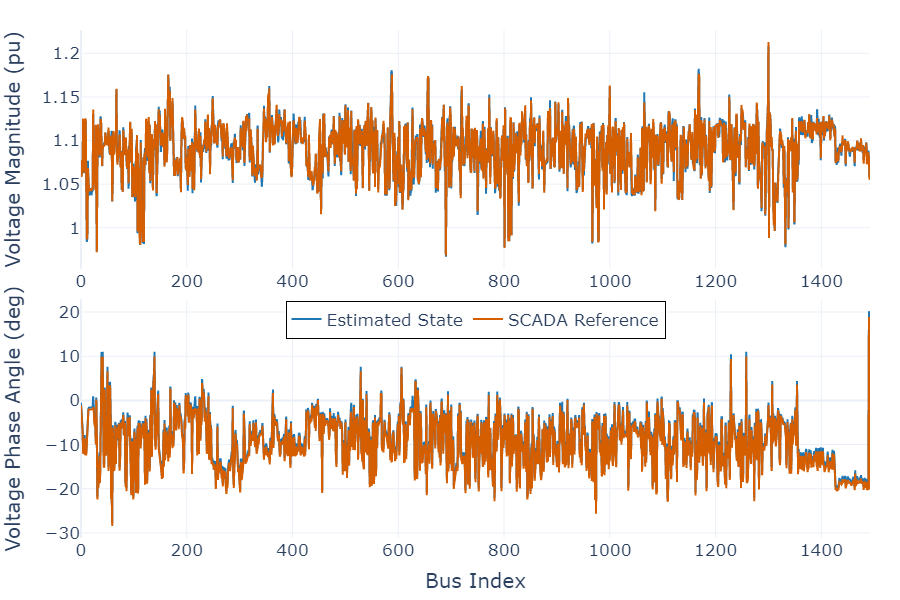}
    \caption{Comparison of DNN-SE$_{VI}$ estimated states with SCADA-based reference values at 2:30~PM.}
    \label{fig:SE_per-bus}
\end{figure}

\subsection{State Estimation Overall Performance on Test System}\label{section:overall-performance}

To assess the overall performance of the developed DNN-SE models (DNN-SE$_{V}$ and DNN-SE$_{VI}$) for System $\mathrm{S}$, we compute the estimation accuracy in terms of mean absolute percentage error (MAPE) for magnitudes and mean absolute error (MAE) for phase angles across all buses in the system. Two complementary evaluations are reported in Table~\ref{tab:SE_results}: a \textit{held-out synthetic test set with known ground truth}, and \textit{evaluation using field PMU data under different load variability conditions}.


To first validate the DNN architecture itself in a setting where ground-truth states are exactly known, the $25{,}000$ synthetic OCs generated through the offline workflow were partitioned into training ($65\%$), validation ($20\%$), and test ($15\%$) subsets, with the test subset withheld entirely from training and drawn from the same distribution as the training data. Because the test subset consists of PF-derived states, exact ground-truth voltage magnitudes and phase angles are available for all $1{,}493$ buses, enabling direct quantitative error analysis. As reported in the first row of Table~\ref{tab:SE_results}, both DNN-SE models achieve very low errors on this evaluation, with DNN-SE$_{VI}$ outperforming DNN-SE$_{V}$ owing to the inclusion of current phasor measurements. This confirms that the DNN architecture itself learns an accurate measurement-to-state mapping.

Evaluation on field PMU data presents a more challenging scenario, as the training data is generated synthetically from historical SCADA snapshots while the test data consists of real PMU streams, introducing a training/test distribution mismatch.
Under the original load variation conditions derived from historical SCADA data, the DNN-SE$_{VI}$ model achieves lower estimation errors than the DNN-SE$_{V}$ model for both voltage magnitudes and phase angles, as listed in the second row of Table~\ref{tab:SE_results}.

While the observed performance is acceptable for a large-scale transmission system, the estimation accuracy is still impacted by the level of load variability present during training.
As illustrated in Fig.~\ref{fig:kde_loads}, the historical SCADA snapshots used for training data exhibit substantial load variability ranging from $\approx$$5$\% to $35$\% across different load buses.
Transmission systems are sensitive to load changes, and small perturbations can produce significant state variations near operational limits or in nonlinear regions of the PF solution space.
To assess the sensitivity of the proposed DNN-SE models to load variation, an additional experiment was conducted in which synthetic load variations with a reduced variation level of $5\%$ were applied around a single SCADA snapshot corresponding to the 2:30~PM operating point.
The load perturbations were generated using a Gaussian distribution centered at the nominal load values.
The estimation results obtained under this reduced load variation scenario are reported in the third row of Table~\ref{tab:SE_results}.
As expected, both models exhibit a significant improvement in estimation accuracy compared to the original load variation case, since the reduced variability leads to more stable OCs and allows the DNN to learn a more consistent mapping between input measurements and system states.

To contextualize these errors against operational requirements, note that the IEEE standard specifies a $1\%$ total vector error (TVE) \parencite{ieee2011c37} which translates to an error of up to $\approx1\%$ in magnitude and up to $0.573^\circ$ in phase angle. Voltage magnitude accuracies for both DNN-SE models comfortably satisfy this threshold across all evaluation settings. For phase angles, the models meet the threshold under reduced ($5\%$) load variability but exceed it under the original load variability, indicating that the diversity of training time OCs is a key factor for operational-grade phase angle accuracy.

\begin{table}[t]
\centering
\small
\caption{State estimation performance under different evaluation settings}
\label{tab:SE_results}
\renewcommand{\arraystretch}{1.2}
\setlength{\tabcolsep}{4pt}
\begin{tabular}{l@{\hskip 6pt}cc@{\hskip 6pt}cc}
\toprule
\multirow{2}{*}{\textbf{Evaluation Setting}}
& \multicolumn{2}{c}{\textbf{DNN-SE$_{V}$}}
& \multicolumn{2}{c}{\textbf{DNN-SE$_{VI}$}} \\
\cmidrule(lr){2-3} \cmidrule(lr){4-5}
& \makecell{\textbf{MAE}\\\textbf{(deg)}} & \makecell{\textbf{MAPE}\\\textbf{(\%)}}
& \makecell{\textbf{MAE}\\\textbf{(deg)}} & \makecell{\textbf{MAPE}\\\textbf{(\%)}} \\
\midrule
Held-out synthetic
& 0.042 & 0.250
& 0.041 & 0.202 \\
Field PMU (original)
& 1.274 & 0.688
& 0.851 & 0.396 \\
Field PMU (5\% variation)
& 0.209 & 0.150
& 0.177 & 0.083 \\
\bottomrule
\end{tabular}
\end{table}

\subsection{Computational Performance and Runtime Analysis}

After high estimation accuracy, high computational speed is the next most critical requirement for time-synchronized PMU-based SE.
To be operationally viable, the proposed estimator must operate within PMU reporting intervals, typically on the order of $\leq$$33~\mathrm{ms}$ for 30~Hz PMU data streams.
In this study, the average online inference time required to produce a single state estimate was evaluated for both the DNN-SE models.
The measured average computation times were approximately $1.5\times10^{-4}$~s for DNN-SE$_V$ and $2.0\times10^{-4}$~s for DNN-SE$_{VI}$, respectively.
These timings correspond to the forward-pass execution time of the trained networks and exclude all offline training computations.
Despite the large scale of System $\mathrm{S}$ with thousands of buses and nearly three thousand state variables, both models consistently operate well within PMU timescale requirements.
The superior computational performance of the DNN-SE approach can be attributed to its non-iterative nature, as state estimates are obtained through a single forward pass of the network.
In contrast to conventional model-based estimators, which require iterative numerical solvers and may exhibit variable execution times and convergence issues, 
the proposed DNN-SE framework was found to provide a more consistent
runtime behavior which is a desirable property for real-time EMS deployment.

\subsection{Bad Data Detection and Replacement (BDDR)}\label{section:BDDR}
All results presented so far assume the use of BDDR during online operation of the DNN-SE framework.
This subsection explicitly examines the significance of BDDR as a preprocessing mechanism and its impact on DNN-SE performance under realistic PMU data conditions.
Unlike the more deterministic PMU data conditioning steps described in Section~\ref{section:online}, the BDDR module is designed to identify and mitigate statistical outliers in real-time and is invoked only when anomalous measurements are detected.
Real-time PMU streams are susceptible to sensor malfunctions, communication errors, cyber-attacks, and transient spikes, which can degrade data-driven estimators that rely on training-time statistical consistency.
To address these challenges, a statistical technique previously proposed in \parencite{varghese2024deep} is employed as a preprocessing step prior to DNN inference. The method leverages the Wald Test to identify potentially faulty or unreliable PMU measurements in real-time.

During the training phase of the DNN-SE, normalized input features corresponding to high-quality data are used, enabling the establishment of expected statistical bounds for each measurement channel. Incoming PMU measurements that violate these bounds are flagged as bad data.
Each input feature of a PMU measurement vector undergoes this test independently and in parallel, allowing timely identification of anomalous values. When bad data are detected, the affected measurements are replaced with the mean value of the corresponding feature computed from the training dataset.
This simple yet effective replacement strategy preserves input dimensionality and ensures compatibility with the trained DNN while preventing corrupted measurements from propagating through the estimation process.

Given the stringent timing requirements of PMU-based SE, it is critical that the BDDR procedure operates within the PMU reporting interval. The proposed BDDR implementation was found
to complete detection and replacement within 33 ms,
ensuring that real-time SE is not delayed and that acceptable
latency is maintained.
The effectiveness of the BDDR approach is illustrated in Fig.~\ref{fig:BDDR_results}, which compares 
voltage magnitude estimation errors at five anonymized PMU-equipped buses before and after applying BDDR.
The results demonstrate a substantial reduction in estimation error following BDDR, highlighting the importance of preprocessing in enabling reliable DNN-SE under real-world data conditions.

\begin{figure}[ht]
    \centering
    \includegraphics[width=0.91\linewidth]{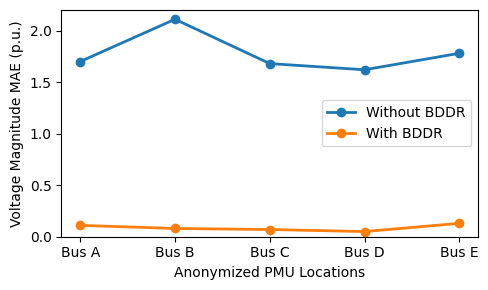}
    \caption{Impact of BDDR on voltage magnitude estimation accuracy at selected anonymized PMU-equipped buses.}
    \label{fig:BDDR_results}
\end{figure}

\section{Conclusion}\label{section:conclusions}
This paper presented the application of an ML--based SE framework, specifically DNN-SE, using actual PMU and historical SCADA data from a large-scale transmission system.
Unlike prior studies that primarily rely on simulated benchmark systems, this work focused on assessing the practical applicability, scalability, and real-time feasibility of data-driven SE under realistic operating constraints, including partial PMU observability and imperfect measurement data.

The results demonstrate that a DNN-based SE model is capable of accurately estimating voltage magnitudes and phase angles across thousands of buses, even in the absence of full system observability.
The study further examined the sensitivity of estimation performance to load variability, revealing that larger variability in the OCs during training vs. testing can significantly impact accuracy.
From an operational perspective, the DNN-SE framework was shown to meet stringent real-time requirements, achieving inference times several orders of magnitude faster than PMU reporting intervals, even for a system with thousands of state variables.
Additionally, the integration of a BDDR preprocessing step was demonstrated to be critical for maintaining estimation accuracy under realistic PMU data quality issues, reinforcing the necessity of robust data handling in real-world deployments.

Several directions are identified for future work, including (a) extending the field PMU validation across topology transitions, (b) incorporating contingency-based and extreme-event scenarios into the training data generation to 
improve robustness and coverage of edge-case system behaviors, and (c) adding flags during actual implementation to let operators know which measurements are identified as bad by the BDDR module.


\vspace{1em}
\noindent\textbf{Acknowledgments}

\noindent This work was supported in part by the Electric Power Research Institute (EPRI) grant 3002029708, the U.S. National Science Foundation (NSF) grant ECCS-2145063, and the U. S. Department of Energy (DOE) grant DE-OE0000983.

The views expressed herein do not necessarily represent the views of the U.S. Department of Energy or the United States Government.

\printbibliography

\endgroup
\end{document}

%% file: hicss-packages.tex
\usepackage[letterpaper]{geometry}
\usepackage{hicss}
\usepackage{times}
\usepackage[none]{hyphenat}
\usepackage{url}
\usepackage{latexsym}
\usepackage{indentfirst}
\usepackage{graphicx}
\graphicspath{{images/}}
\usepackage[
    style=apa,
  ]{biblatex}